\documentclass[sigconf]{acmart}
\usepackage{multirow}
\usepackage{tabularx}
\usepackage{subfigure}
\usepackage{blindtext}
\usepackage{enumitem}
\usepackage{algorithm}  
\usepackage{algpseudocode}  
\usepackage{amsmath}  
\usepackage{hyperref}
\usepackage{flushend}
\usepackage{balance}
\usepackage{mathtools}
\usepackage{bbm}
\usepackage{booktabs}
\usepackage{graphicx}
\usepackage{diagbox}
\usepackage{amsmath}
\usepackage{bm}
\DeclareMathOperator*{\maximize}{\text{\fontfamily{pcr}\selectfont maximize}}

\AtBeginDocument{%
  }

\setcopyright{acmlicensed}
\copyrightyear{2025} 
\acmYear{2025} 

\begin{document}

\title{EGA-V1: Unifying Online Advertising with End-to-End Learning}


\author{Junyan Qiu$^{*\dagger}$, Ze Wang$^{*\dagger}$, Fan Zhang, Zuowu Zheng, Jile Zhu, Jiangke Fan, Teng Zhang,\\ Haitao Wang, Yongkang Wang, Xingxing Wang}

\affiliation{%
 \institution{Meituan, Shanghai, China}
 \city{}
 \country{}
}
\email{{qiujunyan, wangze18, zhangfan133, zhengzuowu, zhujile, jiangke.fan, zhangteng09}@meituan.com}
\email{{wanghaitao13, wangyongkang03, wangxingxing04}@meituan.com}










\renewcommand{\shortauthors}{Junyan Qiu, et al.}

\begin{abstract}
\renewcommand{\thefootnote}{\fnsymbol{footnote}}
\footnotetext[1]{Equal contribution.}
\renewcommand{\thefootnote}{\fnsymbol{footnote}}
\footnotetext[2]{Corresponding authors.}

    Modern industrial advertising systems commonly employ Multi-stage Cascading Architectures (MCA) to balance computational efficiency with ranking accuracy. However, this approach presents two fundamental challenges: (1) performance inconsistencies arising from divergent optimization targets and capability differences between stages, and (2) failure to account for advertisement externalities - the complex interactions between candidate ads during ranking. These limitations ultimately compromise system effectiveness and reduce platform profitability. In this paper, we present \textbf{EGA-V1}, an end-to-end generative architecture that unifies online advertising ranking as one model. EGA-V1 replaces cascaded stages with a single model to directly generate optimal ad sequences from the full candidate ad corpus in location-based services (LBS). The primary challenges associated with this approach stem from high costs of feature processing and computational bottlenecks in modeling externalities of large-scale candidate pools. To address these challenges, EGA-V1 introduces an algorithm and engine co-designed hybrid feature service to decouple user and ad feature processing, reducing latency while preserving expressiveness. To efficiently extract intra- and cross-sequence mutual information, we propose RecFormer with an innovative cluster-attention mechanism as its core architectural component. Furthermore, we propose a bi-stage training strategy that integrates pre-training with reinforcement learning-based post-training to meet sophisticated platform and advertising objectives. Extensive offline evaluations on public benchmarks and large-scale online A/B testing on industrial advertising platform have demonstrated the superior performance of EGA-V1 over state-of-the-art MCAs.
\end{abstract}



\maketitle

\section{INTRODUCTION}
\begin{figure}[htbp]
  \centering
  \includegraphics[width=1\linewidth]{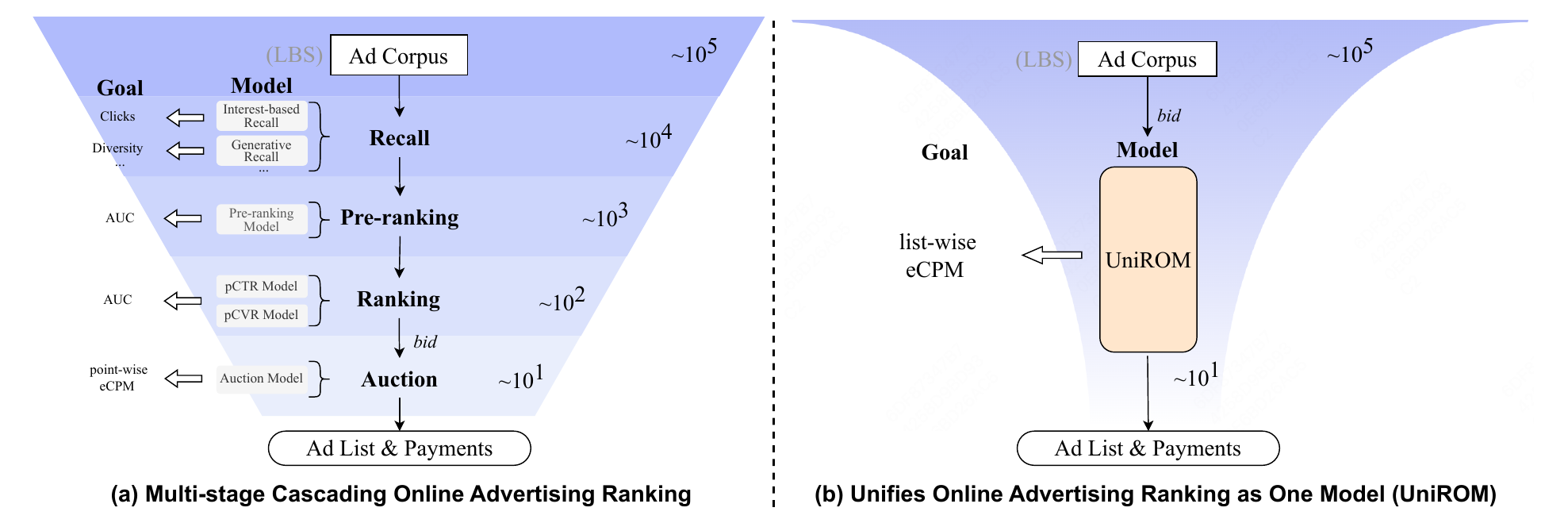}
  \caption{
  Illustrations of multi-stage cascading architecture and unified online advertising ranking.
  }
  \label{fig:overview}
\end{figure}

Online advertising serves as a cost-efficient and precise channel for advertisers to promote contents to millions of online users, which has become the main revenue source for many platforms \cite{edelman2007internet, jansen2008sponsored,liao2022cross}. 
To balance computational efficiency and prediction accuracy, most industrial advertising systems widely adopt the Multi-stage Cascading Architectures (MCAs) \cite{wang2011cascade,covington2016deep,li2022inttower,wang2024adaptive}.
As illustrated in Figure \ref{fig:overview}, a typical MCA decomposes ad ranking problem into four sequential stages: recall, pre-ranking, ranking, and auction. Each stage progressively filters the candidate pool by selecting top-performing ads from its input list before passing them to the subsequent stage.

Although the MCA paradigm has proven efficacy, it is constrained by inconsistencies across stages, which stem from divergent modeling objectives and uneven capacity distribution. These limitations hinder the holistic optimization of platform-wide objectives. For instance, lightweight recall models prioritize speed over accuracy, while subsequent ranking stages employ complex architectures for precise CTR estimation. This misalignment creates prediction discrepancies that propagate through the pipeline, ultimately degrading the quality of final displayed ads \cite{wang2020cold}. Second, ignored externalities—the mutual influence among candidate ads—fundamentally limit performance. Most existing ranking approaches predominantly rely on the independent CTR assumption \cite{li2024deep}, failing to account for how ad permutations or contextual interactions shape user preferences.

To address optimization conflicts in MCAs, the development of coordinated learning frameworks that reconcile inter-stage objectives represents a critical research frontier. Some approaches employ a joint consistency loss computation between the pre-ranking and ranking stages to achieve rank alignment \cite{gu2022ranking,zhang2023rethinking,zhao2023copr}. However, pre-ranking still deals with a much larger pool of candidates than ranking. Relying solely on ranking logs fails to adapt to changes in the retrieval distribution, resulting in a sample selection bias problem \cite{zhao2025hybrid}. Recent advances in multistage optimization have introduced novel paradigm shifts through dynamic sample construction strategies, where each stage operates on specially designed training sets that maintain task-specific characteristics while preserving inter-stage dependency \cite{qin2022rankflow,zheng2024full,zhao2025hybrid}.

Despite these efforts to enhance overall recommendation performance by enabling interaction among rankers, existing approaches predominantly treat each ranker independently, thereby preserving a fragmented architecture where complementary strengths among models remain underutilized. Recent breakthroughs in large language models (LLMs) have catalyzed transformative developments in recommendation systems \cite{zhai2024actions,rajput2024recommender,luo2024qarm,yang2025sparse}, enabling the direct generation of personalized item sequences from user interaction histories. By framing recommendation as sequential transduction tasks, HSTU \cite{zhai2024actions} scales to trillions of parameters and delivers outstanding performance. But it risks vulnerability to cold-start issues and embedding instability due to the dynamic, high-cardinality nature of item IDs. Recently, there emerges a new line of research that indexes items with meaningful IDs using vector quantization algorithms and generate items from the entire item set for recommendation \cite{zheng2024adapting,deng2025onerec,luo2024qarm,yang2025sparse}. Although these methods have the potential to unify cascaded stages into a single model, the inherent sequential dependency in auto-regressive paradigm prohibits parallel computation and results in suboptimal inference latency in online advertising system. 

In this paper, we present \textbf{EGA-V1}, a novel framework that unifies online advertising ranking as one model in location-based services (LBS). By narrowing the candidate set to ads within the same city (reducing its size to $\sim 10^5$), EGA-V1 can perform fine-grained contextual modeling across the entire candidate space, enabling both more accurate item representations and sufficient exploration of user interests. Furthermore, this architecture conceptualizes online advertising systems as a unified generative process, thereby eliminating the inherent goal conflicts between different pipeline stages. Nevertheless, the proposed architecture is confronted with three fundamental challenges: 

\begin{itemize}[leftmargin=*]
\item \textbf{Costly feature processing.} The reliance on complex feature services in online advertising leads to significant storage and transmission overhead, especially for cross-feature interactions between massive user and ad inventories.

\item \textbf{Computational intensity of intra-sequence modeling} Effective advertising requires modeling deep user interests and contextual interactions, but advanced techniques like multi-head attention \cite{vaswani2017attention} face scalability issues due to quadratic computational complexity, making them impractical for large-scale online advertising systems.

\item \textbf{Misalignment with advertising objectives.} Existing generative approaches, primarily developed for recommendations, fail to address critical advertising-specific requirements including auction dynamics, bid optimization, and advertiser objectives.
\end{itemize}

To minimize feature storage and transmission overhead while maintaining expressiveness, we develop a novel Hybrid Feature Service (HFS) that optimizes computational efficiency by decoupling user and ad features. Additionally, by enabling batch processing of multiple candidates in a single pass, HFS significantly amortizes computational and I/O costs, thereby improving system scalability. Furthermore, we present RecFormer, an transformative recommendation framework that leverages an innovative cluster-attention mechanism to simultaneously model users' deep interests and contextual externalities while maintaining computational efficiency. To substantially enhance inference speed without compromising key platform optimization objectives, such as user engagement and revenue performance, EGA-V1 incorporates an AucFormer module that generates advertising sequences through non-autoregressive processing. Finally, we explore a multi-stage training strategy to ensure consistency with platform objectives and auction constraints. To summarize, our main contributions are in three-fold:

\begin{itemize}[leftmargin=*]
    \item We present an innovative End-to-End Generative architecture capable of directly producing ad sequences through a unified model framework. To the best of our knowledge, this is the first industrial-grade solution that comprehensively addresses these challenges.
    
    \item We propose a novel algorithm-engineer co-design framework that integrates hybrid feature service with cluster-attention based modules, facilitating the evolution of advertising systems from feature-centric updates to computation-driven scalability. 

    \item These contributions are rigorously validated through offline experiments on industrial datasets and large-scale A/B tests, demonstrating statistically significant improvements in CTR (+5.2\%), RPM (+13.6\%), and advertiser ROI (+3.1\%) over state-of-the-art MCAs.
\end{itemize}

\begin{figure*}[htbp]
  \centering
  \includegraphics[width=\textwidth]{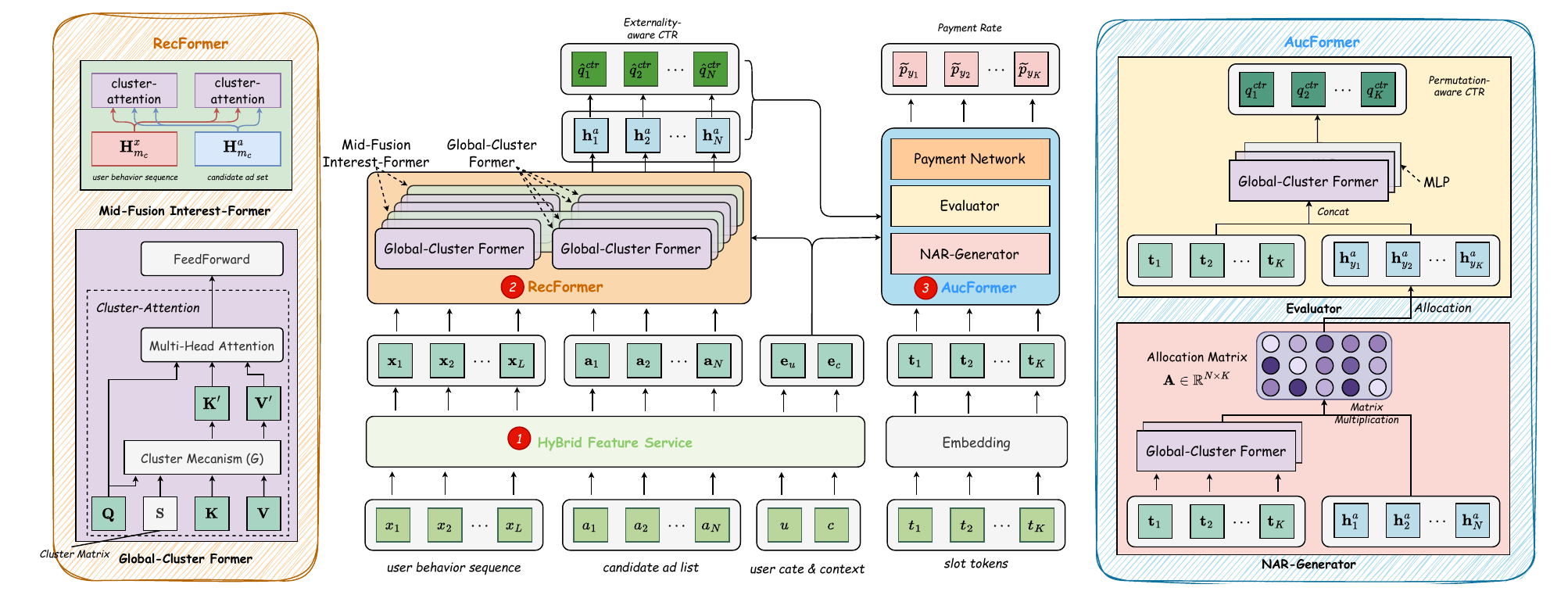}
  \caption{
  Architecture Overview of EGA-V1, showcasing its key components including the Hybrid Feature Service, RecFormer and AucFormer.
  }
  \label{fig:UniROM}
\end{figure*}

 \section{RELATED WORKS}
\textbf{Multi-stage Cascading Architectures} (MCAs) have become the de facto paradigm in industrial advertising systems to balance computational efficiency and prediction accuracy \cite{liu2024recflow, qin2022rankflow}. A typical implementation comprises four stages: recall, pre-ranking, ranking, and auction. The recall stage employs lightweight methods like collaborative filtering \cite{he2017neural} and SASRec \cite{kang2018self} to efficiently retrieve tens of thousands of candidates from the ad corpus. While these approaches enable low-latency computation through simplified architectures, their restricted model capacity inherently limits both feature representation power and ultimate performance ceiling.
In the pre-ranking stage, models such as DSSM \cite{huang2013learning} and COLD \cite{wang2020cold} attempt to approximate ranking model performances under stricter latency constraints. However, the model capacity disparities often leads to prediction discrepancies between pre-ranking and subsequent ranking stages, ultimately compromising system consistency.
The ranking stage utilizes sophisticated architectures like DIN \cite{zhou2018deep} and SIM \cite{pi2020search} with enriched features for precise CTR estimation. Nevertheless, most existing approaches operate under the separable CTR assumption \cite{li2024deep}, failing to account for mutual influences among candidate ads (i.e., externalities) that significantly affect user preferences in real-world scenarios.
The final auction stage implements mechanisms including GSP \cite{edelman2007internet}, DNA \cite{liu2021neural}, and CGA \cite{zhu2024contextual} to allocate ads based on platform revenue objectives. However, the cascaded nature of MCA restricts these mechanisms' capacity to model complete externalities, as early-stage filtering substantially reduces the candidate space before auction occurs.

\section{METHODOLOGY}
As illustrated in Figure \ref{fig:UniROM}, EGA-V1 operates through three integrated components: 1) The Hybrid Feature Service (HFS) efficiently extracting and processing large-scale, fine-grained features; 2) RecFormer modeling deep user interests and candidate externalities across the full candidate set; and 3) AucFormer optimizing ranking to align with advertising platform objectives. Finally, we employ a bi-stage training strategy, wherein the pre-training phase focuses on modeling user preferences, while the post-training phase optimizes for profitability.

\subsection{Preliminaries}
\textbf{Basic task.} 
We describe a typical task in online advertising systems. 
Formally, when a PV (page view) request from the user $u$ arrives, there are $N$ advertisers competing for $K$ ad slots $(K<N)$, denoted as $\mathcal{C} = \{a_1,a_2,\dots,a_N\}$. 
Each advertiser $i$ submits a click bid $b_i$ based on its private click value $v_i$. 
And ad systems output the pCTR denoting the probability that the user clicks the ad.
Given ad auction mechanism $\mathcal{M} \langle \mathcal{A},\mathcal{P} \rangle$, the goal is to propose a winning ad sequence $Y = (a_{y_i} \mid a_{y_i} \in \mathcal{C}, \forall i \in [K])$ that maximizes the expected revenue of ad platform, as follows:
\begin{equation}
  \begin{aligned}
  \maximize_{\theta} \quad & \mathbb{E}_{Y} \Big(
  \sum_{i=1}^K p_{i} \times pCTR_{i} \Big),\\
  \end{aligned}
  \label{eq:problem}
\end{equation}
where $\theta$ is parameter of ad systems and $p_i$ represents the payment of the $i$-th advertiser in ad sequence $Y$.

\textbf{Auction constraints.} Unlike recommender systems, ad systems not only maximize platform revenue but also ensure the utility for advertisers.
Given the auction mechanism $\mathcal{M} \langle \mathcal{A},\mathcal{P} \rangle$, an advertiser's expected utility $u_i$ can be expressed as:
\begin{equation}
  \begin{aligned}
  u_i(v_i;\bm{b}) = (v_i-p_i) \times pCTR_{i}.
  \end{aligned}
\end{equation}
For the design of ad auction mechanisms, two essential properties of ad auction: dominant strategy incentive compatible (DSIC, or IC) and individually rational (IR) are standard economic constraints that must be considered \cite{li2023learning}. 
Specifically, an auction mechanism $\mathcal{M} \langle \mathcal{R},\mathcal{P} \rangle$ is IC, if for each advertiser truthfully reports his bid $b_i = v_i$ and then his utility is maximized. Formally, let $\bm{b}=\{b_1,b_2,\dots,b_N\}$ be the bid profile of all ads, we use $\bm{b}_{-i}$ to represent the bid profile of all ads except ad $i$. For each ad $i$, it holds that
\begin{equation}
u_{i}(v_i;v_i, \bm{b}_{-i}) \ge u_{i}(v_i;b_i, \bm{b}_{-i}), \forall v_i, b_i \in \mathbb{R}^{+},
\end{equation}
and an auction mechanism is IR, if for each advertiser would not be charged more than his bid, as follows:
\begin{equation}
p_i \le b_i, \forall i \in [N].
\end{equation}

\textbf{Problem formulation.} 
The complete problem can be described as developing an architectural solution to successfully select a winning ad sequence with maximum platform revenue $Y$ from a wide range of candidates $X$, while adhering to the constraints of IC and IR. The objective is formulated as follows:

\subsection{Hybrid Feature Service}
\begin{figure}
    \centering
    \includegraphics[width=\linewidth]{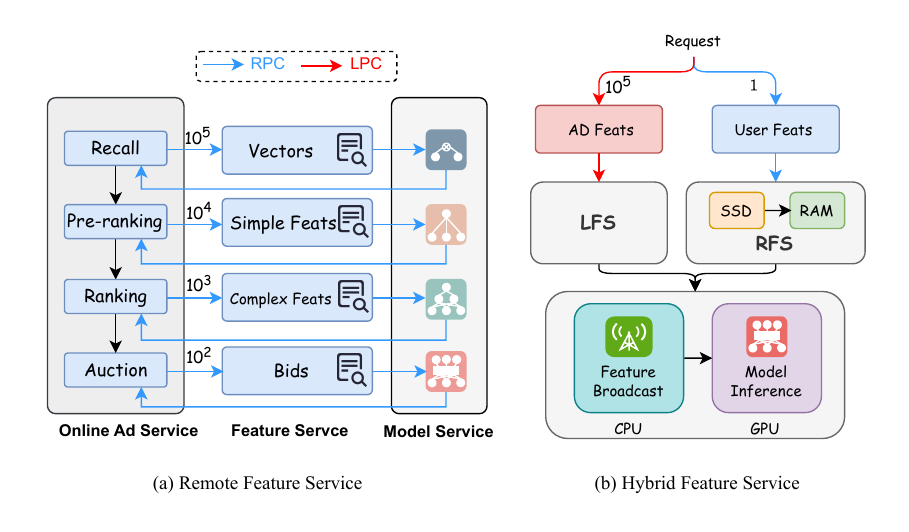}
    \caption{Architectures of two feature services.}
    \label{fig:HFS}
\end{figure}

Traditional Multi-stage Cascading Architectures (MCAs) rely on Remote Feature Services (RFS) to manage high-dimensional feature spaces involving categorical, numerical, cross-modal, and sequential interactions \cite{pi2020search, zhai2024actions}. As illustrated in Figure \ref{fig:HFS} (a), RFS extracts features from distributed storage and transmits them via Remote Procedure Calls (RPC) across stages. While effective for multi-stage coordination, this approach introduces two critical limitations. 
First, the explicit cross-feature space between hundreds of millions of users and tens of millions of candidate ads is too large, creating prohibitive storage and transmission costs when processing all candidate ads per request. This is also a primary reason why advertising systems are constrained to adopt MCA for gradual filtering.
Second, repeated RPC communications between feature services and multiple stages incur substantial latency overhead, particularly when handling high-frequency requests in real-time advertising systems. These inefficiencies fundamentally constrain model expressiveness and system scalability.


To address these challenges, we propose the Hybrid Feature Service (HFS), a computation-aware paradigm that achieves more efficient feature processing through categorized feature storage, as shown in Figure \ref{fig:HFS} (b). The system handles two distinct types of features with different optimization strategies. Candidate ad features, which are numerous in volume, are updated relatively infrequently. To address the network bandwidth challenge posed by $\sim10^5$ candidate ad features, we implement a local storage solution using memory and SSDs, effectively reducing RPC overhead. Conversely, user-side features update at high frequency but only need to be calculated once per request. To accommodate this pattern efficiently, we store these dynamic features in the Remote Feature Service (RFS) and access them through RPC calls, ensuring real-time data availability while maintaining system performance.

For candidate ads, HFS retains only core categorical feature while omitting numerical and explicit cross-features. This simplification reduces the per-ad feature dimensionality and the whole feature space, enabling storage of ad embeddings through Local Feature Service (LFS). The first-order representation of ad $a_i$ is:
\begin{equation}
  \begin{aligned}
  \bm{a}_i = \text{Concat}\Big( \text{Emb}(f_1(a_i)), \text{Emb}(f_2(a_i)), \dots, \text{Emb}(f_{N_f}(x_i)) \Big),
  \end{aligned}
\end{equation}
where $N_f$ denotes the number of preserved categorical features, $\text{Emb}(\cdot)$ denotes the embedding layer and $\text{Concat}(\cdot)$ is the concatenation operation. 
We stack $\bm{x}_i$ together into matrix:
\begin{equation}
  \begin{aligned}
\textbf{E}_{ad} = \big[\bm{a}_1, \bm{a}_2, \dots, \bm{a}_N \big], \quad \textbf{E}_{ad} \in \mathbb{R}^{N \times d}.
  \end{aligned}
\end{equation}
For user modeling, HFS consolidates all user-specific features (categorical attributes and behavior sequences) and context features (e.g., click time, location) into a single RPC call per request. These unified user and context embeddings are then broadcast to all candidate ads, eliminating redundant feature retrievals while maintaining contextual awareness. 
Formally, the representation of user is symbolized as $\bm{x}_{i}$, and the stacked representation of user behavior sequence is expressed as $\mathbf{E}_{{bhvr}} =[\bm{x}_1,\bm{x}_2,\dots, \bm{x}_L] \in \mathbb{R}^{L \times d}$, with $L$ representing the length of sequence. 

Crucially, HFS shifts computational complexity from feature engineering to neural architecture design – omitted numerical and cross-features are implicitly recovered through subsequent ineteration modules rather than explicit storage.

\subsection{RecFormer}
Sequence modeling plays a pivotal role in effectively modeling both user interests and item representations. However, conventional approaches face prohibitive computational complexity that scales superlinearly with sequence length and model depth, creating substantial scalability bottlenecks. To address these limitations, we propose RecFormer, a novel framework featuring: (1) a Global Cluster-Former (GCF) that efficiently models intra-sequence relationships, and (2) a Mid-fusion Interest-Former (MIF) for effective cross-sequence mutual information extraction.


\textbf{Global Cluster-Former} \label{sec:gcf}
Existing approaches for modeling mutual influences among user interaction history or candidate ads typically rely on self-attention mechanisms in Transformer architectures \cite{vaswani2017attention, zhu2024contextual}. While effective for small-scale interactions, directly applying Transformer blocks to industrial-scale advertising scenarios with $L=10^3$ user behavior data and $N=10^5$ candidate ads per request incurs prohibitive computational complexity of $\mathcal{O}(N^2d)$\footnote{$\mathcal{O}(L^2d)$ for user behavior data.}. This quadratic scaling renders conventional Transformers impractical for real-time inference, as the latency and memory overhead become unsustainable. 


To address these challenges, we propose the Global Cluster-Former (GCF) module, which strategically reduces computational complexity while preserving critical externality patterns. GCF consists of $m$ identical layers, each of which can be divided into two sub-layers connected by residual connections \cite{he2016deep} and layer normalization\footnote{Residual connections and layer normalization are omitted for brevity in the following section.}. Take the candidate ads as an example. The foundational and primary sub-layer is the \textbf{cluster-attention}. Concretely, GCF firstly projects input hidden states $\mathbf{H}_{ad}^{\ell}$ ($\ell$ denotes the $\ell_{th}$ layer, particularly, $\mathbf{H}^{0}_{ad}=\mathbf{E}_{ad}$) from previous layer into queries, keys and values $\mathbf{Q,K,V}$. To reduce computational complexity, GCF employs a clustering mechanism $G(\cdot)$ that aggregates the original $N$ keys and values into $N_c$ cluster-level representations. This approach computes attention from the query $\mathbf{Q}$ to these aggregated keys and values, rather than attending to all N candidates directly. Then, the output representation of this sub-layer is obtained by calculating the multi-head attention (MHA) \cite{vaswani2017attention} among the original queries and clustered keys and values.
\begin{gather}
\mathbf{Q,K,V}=Split\Big(\phi(f_1\big(\mathbf{H}_{ad}^{\ell})\big)\Big) \label{eq:split} \ \in\mathbb{R}^{N\times d}\\
\textbf{K}'= G(\textbf{Q}, \textbf{K}, \textbf{S}), \ 
\textbf{V}' = G(\textbf{Q}, \textbf{V}, \textbf{S}), \quad \textbf{K}',\textbf{V}' \in \mathbb{R}^{N_c \times d}\\
\text{head}_i = \text{Attention}(\mathbf{Q},\mathbf{K}',\mathbf{V}')  = \text{Softmax}\left(\frac{\mathbf{Q}\mathbf{K}'^{\top}} {\sqrt{d}}\right)\mathbf{V}' \label{eq:head1}\\
\textbf{H}_{attn}^{\ell} = \phi(f_2(\text{Concat}(\text{head}_1,\text{head}_2,\dots,\text{head}_{N_h}))) \label{eq:att}
\end{gather}
where $f_1$ indicates an MLP that projects $X$ from $d$ to $3d$ hidden nodes, and $Split(\cdot)$ partitions the $3d$-dimensional vector into three distinct components. $\phi$ denotes the Dice activation \cite{zhou2018deep}.

The core innovation of clustering mechanism $G(\cdot)$ lies in the application of learnable cluster matrix $\mathbf{S} \in \mathbb{R}^{N \times N_c}$ \cite{van2024cast}, which dynamically groups $\mathbf{K,V}$ into semantically coherent clusters. The cluster matrix serves as a classifier, functioning as proxies to identify semantically similar keys and values within the embedding space. Formally, we begin with calculating surrogate tokens $\mathbf{A_q},\mathbf{A_k},\mathbf{A_v}$ for $\mathbf{Q,K,V}$ using the cluster matrix. These surrogate tokens are aggregated using adaptive ratios $\phi$, which are computed through a linear transformation of the inner product between queries and the corresponding surrogate key or value tokens.
\begin{gather}
     \mathbf{A}_q = \mathbf{Q}\mathbf{S}^T, \mathbf{A}_k = \mathbf{K}\mathbf{S}^T, \mathbf{A}_v = \mathbf{V}\mathbf{S}^T \quad \in\mathbb{R}^{N_c\times d} \label{eq:g1}\\
     \varphi_k = \sigma\big(f_{k_1}(\mathbf{A}_q\mathbf{A}_{v}^\top)\big),\ \varphi_v = \sigma\big(f_{v_1}(\mathbf{A}_q\mathbf{A}_k^\top)\big) \quad \in\mathbb{R}^{N_c\times 1} \\
     \mathbf{K}'= \varphi_k \odot f_{k_2}(\mathbf{A}_q) + (1 - \varphi_k) \odot f_{k_2}(\mathbf{A}_k) \quad \in\mathbb{R}^{N_c\times d} \\
     \mathbf{V}'= \varphi_v \odot f_{v_2}(\mathbf{A}_q) + (1 - \varphi_v) \odot f_{v_2}(\mathbf{A}_v) \quad \in\mathbb{R}^{N_c\times d} \label{eq:g2}
\end{gather}
where $f_{k_1},f_{v_1}:\mathbb{R}^{N_c\times d}\rightarrow \mathbb{R}^{N_c\times 1}$ and $f_{k_2},f_{v_2}:\mathbb{R}^{N_c\times d}\rightarrow \mathbb{R}^{N_c\times d}$ are neural networks, $\sigma(\cdot)$ denotes sigmoid function, $\odot$ means eletment-wise multiplication.
This design reduces computational complexity from $\mathcal{O}(N^2d)$ to $\mathcal{O}(NN_cd)$, where $N_c\! \ll\! N$.
Moreover, GCF exhibits excellent scalability and can easily enhance model performance by stacking multiple blocks. Finally, GCF generates the second-order representations of all candidate ads candidate ads.

The second sub-layer is a feedforward neural network $f_{out}:\mathbb{R}^{N\times d}\rightarrow\mathbb{R}^{N\times d}$ that processes the output from the cluster-attention sub-layer and generates the final output of the layer.
\begin{equation}
    \mathbf{H}^{\ell+1}_{ad}=f_{out}(\mathbf{H}_{attn}^\ell)\label{eq:out}
\end{equation}
The final representations of candidate ads are extracted from the last layer, i.e., $\mathbf{H}_{ad}=\mathbf{H}_{ad}^{m}$. User behavior sequence $\mathbf{H}_{usr}$ is computed in a similar way through Eq. (\ref{eq:split}$\sim$\ref{eq:out}).


\begin{figure}[htbp]
    \hspace*{-0.8cm}
    \centering
    \includegraphics[width=1.1\linewidth]{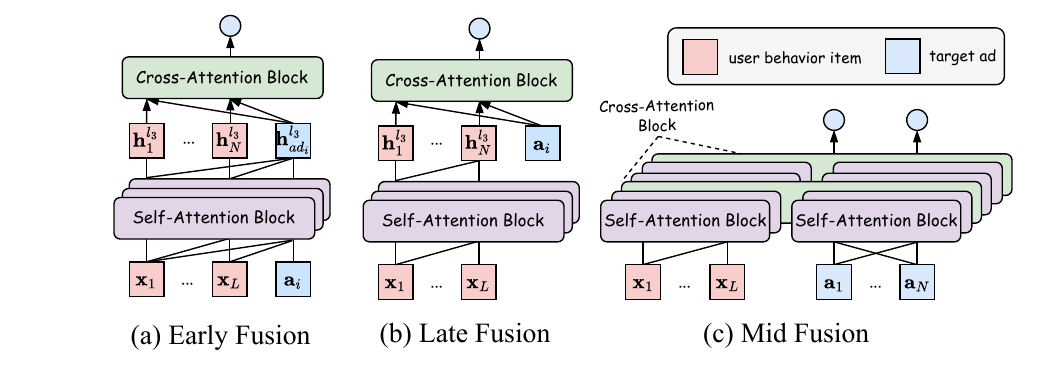}
    \caption{Three paradigms for user interest modeling.}
    \label{fig:fusion}
\end{figure}
\textbf{Mid-fusion Interest-Former}
Existing approaches \cite{zhou2018deep, liao2022deep} for user interest modeling based on user's behavior sequence predominantly adopt two paradigms: \textbf{Early Fusion} and \textbf{Late Fusion}. 
As shown in Figure \ref{fig:fusion}, Early Fusion appends the target ad to the user’s behavior sequence and processes the concatenated sequence through $m$ self-attention blocks. While this paradigm enables full interaction between the target ad and user behaviors, it incurs prohibitive computational complexity of \(O(mN(L+1)^2d)\) due to repeated sequence processing for each candidate ad. Conversely, Late Fusion first extracts user behavior features independently and later interact them with target ad. Although its complexity reduces to \(O((NL+mL^2)d)\), the decoupled interaction limits performance, as user behavior modeling remains agnostic to specific ad during feature extraction. This trade-off between computational efficiency and modeling fidelity poses a critical challenge in large-scale industrial scenarios, where both latency constraints and prediction accuracy are paramount.

To address these limitations, we propose the \textbf{Mid-fusion Interest-Former} (MIF) module that strategically balances interaction granularity and computational overhead. Given the hidden representations of user behavior sequence $\mathbf{H}_{usr}$ and candidate ads $\mathbf{H}_{ad}$ generated by GCF, MIF leverages intermediate hidden states from $m_c$ evenly spaced transformer blocks (e.g., blocks $\{m_c, 2m_c, \ldots, m\}$) to perform cross-interactions. MIF also employs cluster attention except that $\mathbf{Q}$ and $\mathbf{K,V}$ are derived from difference sequences. Conventional approaches employ \textbf{target attention} \cite{zhou2018deep}, where the query derived from the candidate ad attends to the user behavior sequence to evaluate the relevance of target items within the user's historical interactions. Additionally, we propose \textbf{context attention}, which reverses this directional flow by allowing attention signals to propagate from user behaviors to the candidate ad set. This mechanism facilitates capturing deeper user interests within the competitive advertising environment.

Through interval fusion and cluster attention, MIF reduces computational complexity to $O(m_k(N+L)N_cd)$\footnote{$m_k = \left\lceil \frac{m_2}{m_c} \right\rceil$, where $\lceil\cdot\rceil$ denotes the ceiling function that rounds up to the nearest integer. The proof is shown in Appendix \ref{sec:app_A}.}. Assuming that $m=10,m_c=2,N=10^5,L=10^3,N_c=100$, MIF requires only $1/10^4$ the FLOPs of early fusion and half the FLOPs of late fusion.

After obtaining the representations of candidates ads $\textbf{H}_{ad}\in\mathbb{R}^{N\times d}$, MIF incorporates various auxiliary tasks
(e.g., click, purchase) to accelerate representation learning.
Each task is built upon a shared bottom but employs separate parameters for learning. 
Taking the pCTR prediction as an example:
\begin{equation}
 \begin{aligned}
    \hat{q}^{\text{ctr}}_i = \text{MLP}_{\text{ctr}}\big(\text{Concat}(\bm{h}_i, \bm{e}_u)\big),\ \  \forall \bm{h}_i \in \mathbf{H}_{ad}, i\in[1,N],
  \end{aligned}
\end{equation}
where $\hat{q}^{\text{ctr}}_i$ represents the externality-aware pCTR of $x_i$, and $\text{MLP}(\cdot)$ refers to a Multi-Layer Perceptron with a sigmoidal activation in the final layer. The training process details will be discussed in Section \ref{sec:train}.

\subsection{AucFormer}
Traditional auction mechanisms like GSP \cite{edelman2007internet} fail to model externalities effectively and lack deep integration with machine learning technologies. In this paper, EGA-V1 introduces the AucFormer framework, comprising 1) a non-autoregressive (NAR) generator that employs a matching mechanism to simultaneously predict allocation probabilities for candidate ads across multiple interface slots [25], and 2) a permutation-aware evaluator that leverages exposure order information to more accurately estimate user engagement metrics, e.g. pCTR, pCVR.

\textbf{Non-autoregressive generator.} The core idea behind AucFormer lies in using slot tokens as surrogate representations. After processing through several GCF layers to derive slot representations $\textbf{T}\in\mathbb{R}^{K\times d}$, the allocation matrix $\textbf{A}\in\mathbb{R}^{N\times K}$ is computed as:
$$\textbf{A}=\textbf{H}_{ad}\ \textbf{T}^{\top}$$

where $K$ is the number of slots to be assigned. To align the allocation strategy with the platform's profitability objective, we introduce a bid bias to compute the allocation probability for each candidate ad $i$ at slot $k$:
\begin{equation}
  \begin{aligned}
  \label{eq:z}
  \text{z}_{i;k} = \text{Softmax}\Big({ 
  [ e^{w_z} \times \hat q^{\text{ctr}}_j \times b_j + \textbf{A}_{j,k}]_{j=1}^{K}}\Big)_i \ ,
  \end{aligned}
\end{equation}

where $w_z$ represents a parameter that can be learned, with the constraint that $e^{w_z}$ remains positive. This constraint ensures that a higher bid results in a higher probability of allocation.
The IC proof related to the allocation Equation (\ref{eq:z}) follows the derivation theory in CGA. During inference, the generator selects the top $K$ ads ${Y} = [a_{y_1},a_{y_2},\dots,a_{y_K}]$ with the highest scores for each slot. When generating the final winning ad sequence, the generator masks out the ads already assigned to previous slots. For example, when allocating ad for slot $k$, the selected one is determined by $i=\arg\max\limits_{j}(z_{j,k}),j\neq [y_1,y_2,\cdots,y_{k-1}]$.

\textbf{Permutation-aware evaluator.}
The evaluator aims to predict the permutation-aware values for each ad in the ad allocation. Given an ad allocation ${Y} = [a_{y_1},a_{y_2},\dots,a_{y_K}]$, the evaluator combines the corresponding high-order representation $\mathbf{H}_{ad}^{Y} = [\bm{h}_{y_1}, \bm{h}_{y_2}, \dots,\bm{h}_{y_K}]$ with slot embeddings $\mathbf{E}_{t}$, which is then fed into $m_e$ layer cluster-attention based blocks $\mathcal{T}$ to compute the permutation-aware pCTRs .
\begin{equation}
  \begin{aligned}
  q^{ctr}_{y_i} = \text{MLP}_{\text{E}}\Big( 
  \mathcal{T}\big(\text{Concat}(\bm{h}^a_{y_i}, \bm{t}_i, \bm{e}_{u}) \big)
  \Big), \forall i \in [K],
  \end{aligned}
\end{equation}
where $\mathcal{T}(\cdot)$ is determined by the equations (\ref{eq:split}) through (\ref{eq:g2}).

\textbf{Payment network.}
Motivated by the successful application of neural networks to payment in \cite{zhu2024contextual}, AucFormer introduces a payment network to learn the optimal payment rule. Specifically, to satisfy IR constraint, the payment network employs a sigmoidal activation function to compute the payment rate $\tilde{\bm{p}} \in [0, 1]^{K}$, and subsequently outputs the payment $\bm{p} = \tilde{\bm{p}} \odot \bm{b}$. The payment rate of $x_{y_i}$ in $Y$ is:
\begin{equation}
  \begin{aligned}
  \tilde{{p}}_{y_i} = \text{MLP}_{\text{pay}}\Big( \text{Concat}\big(\bm{h}^{a}_{y_i}, q^{\text{ctr}}_{y_i}, \bm{b}_{-y_i}\big) \Big), \forall i \in [K],
  \end{aligned}
\end{equation}
where $\bm{h}^{a}_{y_i}$ represents the high-order representation of $i$-th ad in allocation $Y$, $q^{\text{ctr}}_{y_i}$ denotes the permutation-aware pCTR of $i$-th ad in $Y$, and $\bm{b}_{-{y_i}} \in \mathbb{R}^{K-1}$ is the bids vector of ads in $Y$ excluding the $i$-th ad.
It is evident that the payment network operates independently of the evaluator. During inference, only the generator and the payment network are deployed to reduce computation costs.

\section{TRAINING AND OPTIMIZATION}
\label{sec:train}
In this section, we present a two-stage training framework inspired by large language model optimization \cite{ouyang2022training}. The first stage, pre-training, focuses on aligning EGA-V1 with user engagement signals such as clicks and purchases. Subsequently, the post-training stage employs reinforcement learning to optimize EGA-V1’s response to auction feedback and guide the payment network under predefined economic constraints.

\subsection{Pre-training}
We treat each request as a sample $\kappa$ to build the pre-training dataset $\mathcal{D}$, rather than considering each exposure as a sample.
Each sample $\kappa$ contains $N_s$ unexposed ads selected via popularity sampling \cite{mikolov2013distributed} from the valid candidate ad pool, combined with $K$ exposed ads from actual impressions. The popularity sampling strategy prioritizes frequently occurring candidates while maintaining long-tail coverage through probabilistic selection, thereby enhancing training efficiency without sacrificing representation diversity.

In each sample $\kappa$, we utilize the user's clicks under current request as labels for the $K$ exposed ads and the user's overall click behaviors across the platform as labels for $N_s$ unexposed ads, denoted as $\zeta_i^{clk} \in \{0,1\}$.
The overall click behaviors on one specific ad include the user's clicks from other entries on the current platform as well as the user's clicks when the ad is displayed in its organic item form in one same session.
The pre-training objective minimizes binary cross-entropy loss over both the set-aware pCTR:
\begin{equation}
  \begin{aligned}
\mathcal{L}_{{pt}} = -\frac{1}{|\mathcal{D}|} \!\sum_{\kappa \in \mathcal{D}}\!\!\! \sum_{i=1}^{K + N_s}\!\! \Big(\zeta_i^{clk} \log(\hat q^{\text{ctr}}_i) + (1 - \zeta_i^{clk}) \log(1 - \hat q^{\text{ctr}}_i) \Big).
  \end{aligned}
\end{equation}

For brevity, we only utilize click signals for illustration. Additionally, other user behavioral signals, like purchases, reviews, and so forth, can also contribute to the representation learning through pre-training in a similar manner.

\subsection{Post-training}
During the post-training phase, reinforcement learning is primarily used to fine-tune the model in alignment with the platform's profitability objectives. This process involves three key components: training the reward model, reinforcement learning from auction feedback, and optimizing the payment network.

\textbf{Training the Reward Model}
The permutation-aware evaluator in AucFormer serves as a reward model, designed to evaluate the quality of the generated ad sequence.
To improve training effectiveness and prevent unnecessary repetition, we utilize the pre-trained model's parameters and freeze the parameters of the modules before RecFormer during the reward model training.
Formally, we only use the $K$ exposed ads from each sample $\kappa$ to train the reward model and the loss is calculated as:
\begin{equation}
  \begin{aligned}
\mathcal{L}_{{rm}} = -\frac{1}{|\mathcal{D}|} \!\sum_{\kappa \in \mathcal{D}}\!\sum_{i=1}^{K}\! \Big(\zeta_i^{\kappa} \log(q_i^{ctr}) + (1 - \zeta_i^{\kappa}) \log(1 - q_i^{ctr}) \Big).
  \end{aligned}
\end{equation}

\textbf{Reinforcement Learning from Auction Feedback.}
Following reward model convergence, the non-autoregressive generator undergoes optimization via Reinforcement Learning from Auction Feedback (RLAF) with frozen evaluator parameters.
The reward signal $r_{y_i}$ for ad $a_{y_i}$ in sequence $Y = [a_{y_1}, a_{y_2}, ..., a_{y_K}]$ quantifies its marginal contribution to platform revenue:
\begin{equation}
  \begin{aligned}
r_{{y_i}} = \sum_{a_{y_i} \in Y} b_{{y_i}} q^{ctr}_{{y_i}} - \sum_{a_{y_j} \in Y_{-i}} \tilde{b}_{y_j} \tilde{q}^{ctr}_{y_j}, \quad \forall i \in [K],
  \end{aligned}
\end{equation}
where $Y_{-i}$ represents the best ad sequence excluding $a_{y_i}$, and $\tilde{b}_{y_j} \tilde{q}^{ctr}_{y_j}$ indicates the corresponding bid and pctr. The policy gradient objective maximizes expected rewards through:
\begin{equation}
  \begin{aligned}
\mathcal{L}_{{rlaf}} = -\frac{1}{|\mathcal{D}|} \sum_{\kappa \in \mathcal{D}} \sum_{Y \in \mathcal{X}(\kappa)} \sum_{i=1}^{K} \Big( r_{y_i} \log (z_{y_i;i}) \Big),
  \end{aligned}
\end{equation}
where $\mathcal{X}(\kappa)$ denotes the valid candidate ads in sample $\kappa$ and $Y$ denotes the ad sequence output by the non-autoregressive generator. 
As the needs of advertisers and business goals constantly evolve, we decided to freeze the parameters of RecFormer during RLAF, focusing solely on optimizing the parameters of AucFormer that are closely tied to the business objectives.

\textbf{Optimization of Payment Network.}
The payment network optimization follows generator-evaluator training, employing a Lagrangian dual formulation to balance revenue maximization and IC constraints \cite{zhu2024contextual}. The loss function incorporates both platform revenue and ex-post regret minimization:
\begin{equation}
  \begin{aligned}
\!\mathcal{L}_{{pay}}\! =\! -\frac{1}{|\mathcal{D}|} \!\!\sum_{\kappa \in D} \sum_{Y \!\in\! \mathcal{X}\!(\kappa)}\!\! \left( 
\sum_{x_{y_i}\!\in Y}\! \!p^{Y}_{y_i} q^{ctr}_{y_i}
\!\!-\!\!\!\! \sum_{x_{y_i}\!\in Y}\!\! \lambda_{y_i} \widehat{\text{tgt}}_{y_i}
\!\!-\! \frac{\rho}{2} \!\!\!\sum_{x_{y_i}\!\in Y}\! \!(\widehat{\text{tgt}}_{\!y_i})^2 \right),
  \end{aligned}
  \label{eq:pay_loss}
\end{equation}
where $\widehat{\text{tgt}}_{y_i}$ is the \textit{ex-post regret} for ad $x_{y_i}$ in $Y$, $\lambda_{y_i}$ is the Lagrange multiplier used to balance revenue and IC constraints and $\rho$ is the hyperparameter for the IC penalty term.

\section{EXPERIMENTS}
\label{sec:exp}

In this section, we evaluate the effectiveness of EGA-V1 using offline experiments and online A/B testing.

\subsection{Experiment Setup}
\subsubsection{Dataset}
We provide empirical evidence for the effectiveness of EGA-V1 on the industrial dataset \textbf{Meituan} \footnote{Due to the need to protect business secrets, some transformations were applied to the results. These transformations were carefully designed to maintain the statistical properties while ensuring that no sensitive business - related information could be reverse - engineered from the transformed results.} from a well-known local life platform that leverages Location-Based Services (LBS) to provide users with convenient and tailored services. The industrial dataset is constructed from the real logs of advertising platform from April 2024 to October 2024. It contains 200 million requests from more than 2 million users across nearly 10 million ads. The first 200 days are used for pre-training, a randomly sampled 50 days subset is employed fro post-training, and the last 14 days are used as the test set. In the LBS scenario, the maximum number of valid candidate ads in each sample is $N=100,000$. To enhance training efficiency, popularity sampling is utilized during training, with $N_s=2995$ and $K=5$. During testing, inference is conducted using all candidate ads for each sample, where $N=100,000$ and $K=5$ are employed.


\subsubsection{Evaluation Metrics}
In offline experiments, we utilize standard ranking metrics including Recall@50 and Area Under the Curve (AUC) to evaluate the pre-ranking and ranking stages, respectively. We further employ an offline replay system with the permutation-aware evaluator to evaluate the expected CTR and the expected revenue of the results generated by different architectures.
\begin{itemize}[leftmargin=2em]
    \item $
    \text{eCTR} = \frac{1}{|\mathcal{D}_{\text{test}}|} \sum_{\kappa \in \mathcal{D}_{\text{test}}} \sum_{Y \in \mathcal{X}(\kappa)} \sum_{x_{y_i} \in Y} q^{ctr}_{y_i} \times 100\%.
    $
    \item $
    \text{eRPM} = \frac{1}{|\mathcal{D}_{\text{test}}|} \sum_{\kappa \in \mathcal{D}_{\text{test}}} \sum_{Y \in \mathcal{X}(\kappa)} \sum_{x_{y_i} \in Y} \big(q^{ctr}_{y_i}*p_{y_i}\big) \times 1000.
    $
    \item IC Metric: $\Psi = \frac{1}{|\mathcal{D}_{\text{test}}|} \sum_{\kappa \in \mathcal{D}_{\text{test}}} \sum_{i \in k} \frac{\widehat{\text{tgt}}_i^\kappa}{u_i(v_i^\kappa; \bm{b}^\kappa)}$,
    where $\widehat{\text{tgt}}_i^d$ denotes the empirical ex-post regret for advertiser $i$ in session data $\kappa$, and $u_i$ is the realized utility. This metric evaluates IC, representing the relative utility gain an advertiser could obtain by manipulating its bid~\cite{liu2021neural}. Following~\cite{zhu2024contextual}, IC is empirically tested via counterfactual perturbation: for each advertiser, the bid $b_i$ is replaced with $\gamma \times b_i$, where $\gamma \in \{0.2 \times j \mid j=1,2,\dots,10\}$.
\end{itemize}
In online A/B testing, we introduce the following three metrics to measure platform revenue, user experience, and Return on Investment of advertisers, respectively.
\begin{itemize}[leftmargin=2em]
    \item \textbf{Revenue Per Mille}: $\text{RPM} = \frac{\sum \text{click} \times \text{payment}}{\sum \text{impression}} \times 1000.$
    \item \textbf{Click-Through Rate}: $\text{CTR} = \frac{\sum \text{click}}{\sum \text{impression}}\times 100\%.$
    \item \textbf{Return on Investment}: $\text{ROI} = \frac{\sum \text{gross merchandise volume}}{\sum \text{payment}}.$
\end{itemize}

\subsubsection{Baselines}
We compare EGA-V1 with the following two representative architectures, which are widely used in industry: 1) \textbf{MCA}. The Multi-stage Cascading Architecture is a common design for online advertising systems. To implement this architecture effectively, we employ four representation methods: SASREC \cite{kang2018self} for recall, DSSM \cite{huang2013learning} for pre-ranking, DIN \cite{zhou2018deep} for ranking, and GSP \cite{edelman2007internet} for auction. 2) \textbf{FS-LTR}. Full Stage Learning to Rank \cite{zheng2024full} is a unified training framework designed for multi-stage recommendation systems. It leverages relabeled data from all stages to train models in MCA, ensuring that the top-ranked items are more likely to pass through subsequent stages and align with user interests. To ensure effectiveness and comparability of FS-LTR, we use the same architecture as MCA, incorporating FS-LTR's sample re-labeling technique during training to enhance consistency.

\subsubsection{Hyperparameters} 
For MCA and FS-LTR, we follows the hyperparameters setting in \cite{liu2024recflow}.
For EGA-V1, we tried different hyperparameters using grid search. Due to space limitations, only the most optimal parameters are presented in this paper. 
The clustering approach $G$ employs an adaptive clustering mechanism, the hidden layers of the MLP are 128 and 32, the learning rate is $10^{-3}$, the batch size is $128$ and the optimizer is Adam. the user behavior sequence length $L=1000$, the dimension size $d=128$, the layer number of GCF and MIF $m=6$, the layer number of RecFormer $m_e=3$, the number of interval layers in MIF $m_k=2$, the number of attention heads $N_h=4$ and the number of adaptive clusters $N_c=128$. 

\subsection{Offline Experiments}
Table \ref{tab:result} summarizes the results on industrial dataset from the offline experiments. Each offline experiment is repeated $5$ times with different random seeds and each result is presented in the form of mean ± standard.  
The experimental results yield the following observations. On the industrial dataset, EGA-V1 improves over the state-of-the-art baselines in AUC, Recall@50, eCTR, eRPM, and $\Psi$ respectively, demonstrating the superiority of EGA-V1. Specifically, EGA-V1 achieves a Recall@50 of 0.513, representing a significant lift of 20.4\%\footnote{Lift percentage means the improvement of EGA-V1 over the best baselines} over the strongest baseline FS-LTR. The AUC increases to 0.754 (+1.48\%), and eCTR improves +8.3\%, demonstrating superior user modeling and ranking capabilities. In terms of platform revenue, EGA-V1 achieves an eRPM of 217.1, which is 11.4\% higher than FS-LTR. Most notably, the IC metric $\Psi$ is dramatically reduced to 2.3\%, compared to 9.1\% and 9.4\% for FS-LTR and MCA respectively, indicating much stronger incentive compatibility and robustness to auction constraints.
\begin{table}[h]
  \renewcommand\arraystretch{1}
  \centering
  \caption{The experimental results of different architectures.}
  \setlength{\tabcolsep}{1mm}{
  \begin{tabular}{c|ccccc}
  \toprule
  Method & Recall@50 & AUC & eCTR(\%) & eRPM & $\Psi$ \\
  \midrule
  MCA   & 0.289 & 0.741 & 6.043 & 185.2 & 9.4\% \\
  FS-LTR & 0.426 & 0.743 & 6.140 & 194.9 & 9.1\% \\
  {EGA-V1} & \textbf{0.513}  & \textbf{0.754}  & \textbf{6.652}  & \textbf{217.1} & \textbf{2.3\%} \\
  \bottomrule
  \end{tabular}}
  \label{tab:result}
\end{table}

\subsection{Ablation Study}
To verify the effectiveness of EGA-V1’s various design considerations, we construct three variants:
\begin{itemize}[leftmargin=2em]
    \item $\textbf{EGA-V1}_{-{gcf}}$ removes the GCF and uses point-wise pCTR instead of set-aware pCTR for next tasks.
    \item $\textbf{EGA-V1}_{-{mif}}$ removes the MIF and only uses DIN to capture user's interests from user's historical behavior sequence.
    \item $\textbf{EGA-V1}_{-auf}$ removes the AucFormer and uses GSP mechanism for completing ad allocation and payment.
\end{itemize}

Judging from the online experimental results in Table \ref{tab:ablation}, we have the following findings: 
1) The variant without GCF performs worse than EGA-V1. This phenomenon proves that our proposed GCF can effectively model the global set-aware externalities and help EGA-V1 to achieve better performance. Especially in our LBS scenarios, the benefits of this global externalities modeling are further enhanced.
2) The experimental results of $\textbf{EGA-V1}_{-{mif}}$ are worse than EGA-V1, this supports that the deep modeling of user behavior sequences is crucial for both prediction and generation tasks.
(3) The performance gap between w/ and w/o AucFormer is obvious. This indicates that AucFormer structure and RLAF training approach can effectively align with auction preferences, resulting in improved ad sequence generation and overall performance.

\begin{table}[htbp]
\centering
\caption{Performance comparison of different variants.}
\label{tab:cga_ablation}
\begin{tabular}{l|cc}
\toprule
\textbf{Method} & \textbf{eCTR}(\%) & \textbf{eRPM} \\
\midrule
\textbf{EGA-V1} & \textbf{6.652} &  \textbf{217.1} \\
EGA-V1-$gcf$ &  6.466  (-1.3\%)  &  214.8  (-1.1\%)\\
EGA-V1-$mif$  & 6.389  (-2.5\%)  &  212.5  (-2.1\%)\\
EGA-V1-$auf$  & 6.415  (-2.1\%)  &  208.0  (-4.2\%)\\
\bottomrule
\end{tabular}
\label{tab:ablation}
\end{table}

\subsection{Cross Features Analysis}

This section demonstrates how target-attention (TA) and context-attention (CA) mechanisms in MIF module mitigate performance loss from cross-feature elimination. Furthermore, we also compare the results with the Late Fusion (LF), where the MIF module is replaced by a late fusion approach. In our experiment, we remove 18 cross features (CF) out of the initial 94 features, and the corresponding results are shown in Table \ref{tab:cross}. 
\begin{table}[htbp]
\centering
\caption{Impact analysis of discarding cross-features on AUC and Deviation. Values in the parentheses of the first block show the difference from the underlined value, and the second block shows the difference from the bold value. $\downarrow$ indicates that smaller is better.}
\label{tab:cross}
\begin{tabular}{l|cc}
\toprule
\textbf{Method} & \textbf{AUC} & \textbf{Dev.}$\downarrow $\\
\midrule
MCAs & \underline{0.741} & \underline{0.112} \\
MCAs w/o CF & 0.733 (-0.8\%) & 0.131 (+1.9\%) \\
\midrule
EGA-V1 & \textbf{0.754} & \textbf{0.005} \\
EGA-V1 w/o CF & 0.751 (-0.3\%) & 0.011 (+0.6\%) \\
EGA-V1$_{\text{-CA}}$ w/o CF & 0.745 (-0.9\%) & 0.036 (+3.1\%) \\
EGA-V1$_{\text{-TA}}$ w/o CF & 0.743(-1.1\%) & 0.047 (+4.2\%) \\
EGA-V1$_{\text{-mif}}$ w/o CF & 0.737 (-1.7\%) & 0.048 (+4.4\%) \\
EGA-V1$_{\text{LF}}$ w/o CF & 0.741 (-1.3\%) & 0.037 (+3.2\%) \\
\bottomrule
\end{tabular}
\end{table}

As illustrated in the table, cross-features demonstrate a significant impact on both order prediction (AUC) and accuracy (Dev.)\footnote{Dev. means deviation, which measures the discrepancy between predicted click-through rate (pCTR) and actual click-through rate (CTR). Formally, $\text{Dev.}=|1-\dfrac{\text{pCTR}}{\text{CTR}}|$} performance. Notably, their influence on MCAs is substantially greater than on EGA-V1. When cross-features are omitted, the removal of both context-attention and target-attention mechanisms leads to considerable AUC degradation, with reductions of 0.9\% and 1.1\%, respectively. More critically, removing both components (MIF) results in performance deterioration that nearly equals the combined effect of their individual removals, indicating that these features contribute synergistically rather than independently to model performance. Moreover, replacing MIF with late fusion degrades the AUC by 1.0\% and increases the deviation by 2.6\%. The empirical results demonstrate that MIF robustly addresses the performance drop attributable to cross-feature information loss.

\subsection{Scaling Law}
In EGA-V1, each module uses stacked blocks based on cluster-attention as core structure to ensure scalability. Here, we vary the number of blocks across modules to study scaling laws. Besides, we Results in Figure \ref{fig:m1_2_3} reveal:
\begin{figure}[htbp]
  \centering
  \includegraphics[width=0.9\linewidth]{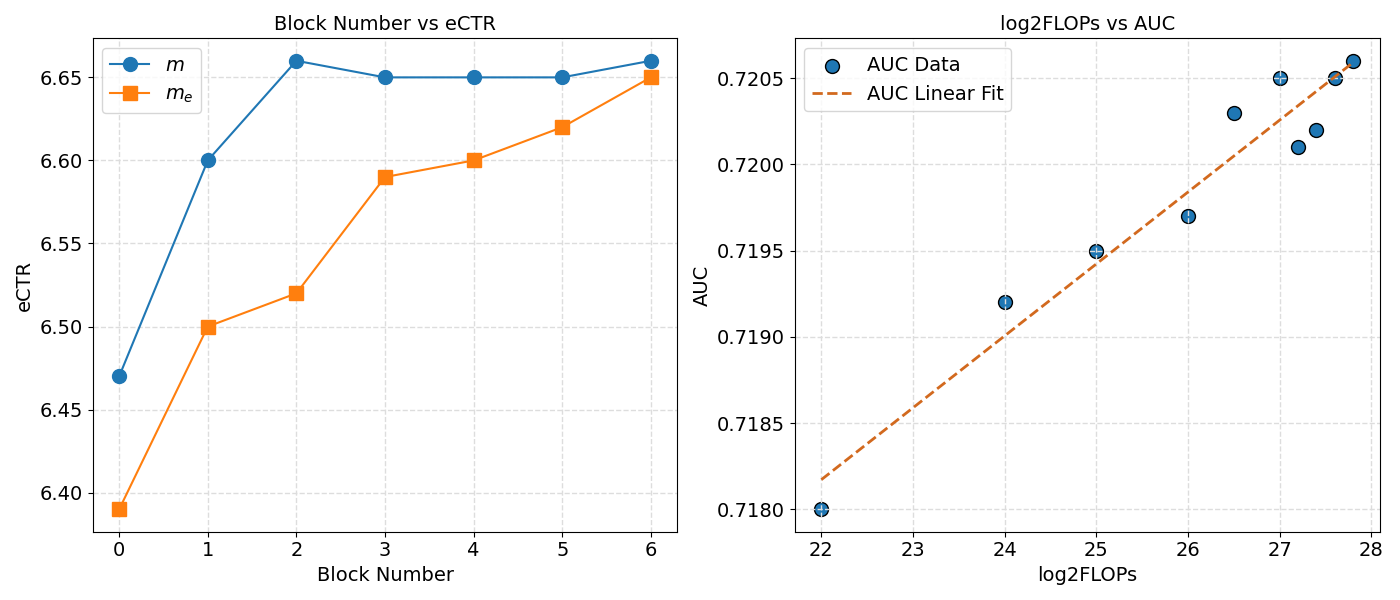}
  \caption{
    Scaling trends featuring the relationship between model layer depth and FLOPs.
  }
  \label{fig:m1_2_3}
\end{figure}

1) The performance curve of and $m$ and $m_e$ illustrates that increasing the number of blocks in RecFormer and AucFormer yields progressive performance gains, though with diminishing returns beyond a critical threshold. This plateau effect likely stems from inherent constraints imposed by fixed data scales and sequence lengths in our experimental setup. While exploring larger datasets and extended sequences could unlock additional scaling potential, such investigations remain beyond the scope of this study due to resource constraints.

2) The figure on the right reveals a clear positive correlation between computational cost and model performance in EGA-V1. The upward-trending linear fit (brown dashed line) indicates that increased computational resources consistently enhance recommendation accuracy, though the marginal gains gradually diminish at higher FLOPs levels. This suggests that while scaling computation remains an effective strategy for performance improvement within the tested range, the model's architecture or data constraints may eventually limit further gains—mirroring the saturation pattern observed in block stacking. The relationship underscores the importance of balancing computational investment against diminishing returns in practical deployment scenarios.

\subsection{Online Results}
\label{sec:online}
To verify EGA-V1’s effectiveness in the real-world, we compare EGA-V1 with the fully deployed MCA in industrial advertising system through online A/B tests. 
Table \ref{tab:online_results} presents the results of online A/B testing conducted from November 18 to November 24, 2024 on a low-traffic ad slot of Meituan. 
Experimental results demonstrate 5.2\% improvements in CTR, 13.6\%  in RPM, and 3.1\% in ROI respectively, indicating EGA-V1's effectiveness in enhancing user experience, platform revenue, and advertiser utility. Notably, the harmonized gains in user engagement and advertiser's ROI suggest that revenue growth stems not from payment inflation, but from enhanced modeling capabilities.

We also analyzed the online inference computational complexity of EGA-V1 and MCA. EGA-V1 processes hundreds of times more candidate ads than MCA's ranking model with only a 5 ms average increase (2.2\% relatively) in online response time (RT) per request). The complexity is discussed in Appendix \ref{sec:app_A}.
\begin{table}[h]
\centering
\caption{Experimental results from Online A/B tests.}
\label{tab:online_results}
\begin{tabular}{c|c|c|c|c}
\hline
{Relative change in metrics} & CTR & RPM & ROI & RT \\ 
\hline
\hline
\textbf{EGA-V1} over baseline-MCA  & +5.2\% & +13.6\% & +3.1\% & +2.2\% \\ 
\hline
\end{tabular}
\end{table}

Recent efforts like FS-LTR \cite{zheng2024full} and COPR \cite{zhao2023copr} attempt to enhance stage consistency through unified training with relabeled multi-stage data. While demonstrating improved performance, such methods remain constrained by the MCA paradigm's structural limitations: 1) The inherent conflict between stage-specific optimization objectives creates irreducible prediction inconsistencies; 2) The sequential filtering mechanism prevents holistic externality modeling across all candidates; 3) Feature engineering dominance limits scalability compared to computation-driven architectures \cite{zhai2024actions}.
Notably, \citet{zheng2024full} developed an enhanced ranking principle to mitigate selection bias in downstream stages, representing the state-of-the-art MCA improvement. However, their solution still requires complex multi-stage coordination rather than offering true end-to-end optimization. These fundamental limitations motivate our investigation into alternative architectural paradigms.

\section{CONCLUSIONS}
In this paper, we propose EGA-V1 that Unifies the entire ad ranking and allocation pipeline as One Model. Our key innovations include: 1) a hybrid feature service that decouples user and ad feature processing to reduce latency while preserving expressiveness; 2) an transformative recommendation framework that leverages an innovative cluster-attention mechanism to efficiently model users’ deep interests and contextual externalities; and 3) a bi-stage training strategy align with user preference and platform objectives. 
Extensive experiments on public and industrial datasets demonstrate EGA-V1's superiority over state-of-the-art approaches.
In the future, we will further explore scaling laws and different training paradigms to enhance EGA-V1's capabilities.

\newpage
\appendix
\section{APPENDIX}
\subsection{Complexity Analysis}
\label{sec:app_A}
We provide a detailed comparison of computational complexity between MCA and EGA-V1. All floating-point operations (FLOPs) are calculated based on standard Transformer block operations \cite{vaswani2017attention}. 
Given sequence length $L$, hidden dimension $d$, the total FLOPs of a standard Transformer block consist of three components:

\begin{equation}
\begin{aligned}
\text{FLOPs}_{block} &= \underbrace{4\times2Ld^2}_{\text{matrix calculation}}+\underbrace{16Ld^2}_{\text{feed-forward network}}+\underbrace{4L^2d}_{\text{attention}}\\
&=4L^2d+24Ld^2.\\    
\end{aligned}
\end{equation}
Similarly, when we change the query to $\mathbf{Q} \in \mathbb{R}^{L_1 \times d}$ and key,value to $\mathbf{K},\mathbf{V} \in \mathbb{R}^{L_2 \times d}$, the total FLOPs of the Transformer block is: 
\begin{equation}
\text{FLOPs}_{block_2} = 4L_1L_2d + 20L_1d^2 + 4L_2d^2.
\end{equation}

\noindent\textbf{MCA Complexity.}
The FLOPs of MCA mainly has two parts:
\begin{enumerate}[leftmargin=*]
    \item \textbf{Pre-ranking Stage}: Using the Dual-tower model \cite{huang2013learning} as a reference, the primary operations involve the computation of the user embedding ($L \times d \times d \times 2$) and the execution of the inner dot product for $N$ candidate ads ($N \times d$):
    \begin{equation}
        \text{FLOPs}_{{pre}} = N d + 2 L d^2.
    \end{equation}
    
    \item \textbf{Ranking Stage}: Taking a Transformer-based model for instance, we use Early Fusion paradigm to compensate for the efficiency decline in MCA. The Flops is:
    \begin{equation}
        \text{FLOPs}_{{rank}} = m_r \big[4N_r(L+1)^2d + 24N_r(L+1)d^2\big].
    \end{equation}
    where $m_r$ is the number of blocks, $N_r$ is the number of valid candidate ads in ranking stage, where $N_r<N$.
\end{enumerate}
Finally, the total Flops of MCA is calcalated as:
\begin{equation}
    \begin{aligned}
    \text{FLOPs}_{mca} =& \text{FLOPs}_{{pre}} +  \text{FLOPs}_{{rank}}.
    \end{aligned}
\end{equation}

\noindent\textbf{EGA-V1 Complexity.}
The FLOPs of EGA-V1 contains three modules:
\begin{enumerate}[leftmargin=*]
    \item \textbf{Global-Cluster Former Module} ($m$ blocks):
    \begin{equation}
        \text{FLOPs}_{{gcf}} = m(6NN_cd + 4NN_cd + 24Nd^2).
    \end{equation}
    where $N_c$ is the number of surrogate tokens and $N_c \! \ll\! N$, $6NN_cd$ is the Flops of cluster mechanism $G$ and $24Nd^2$ is the Flops of origin matrix calculation. 
    
    \item \textbf{Mid-Interest Interest-Former Module} ($m_k$ blocks):
    \begin{equation}
    \begin{aligned}
        FLOPs=&\underbrace{m_kNN_cd}_{\text{target attention}}+\underbrace{m_kLN_cd}_{\text{context attention}}
        \\=&m_k(N+L)N_cd
    \end{aligned}
    \end{equation}
    
    \item \textbf{AucFormer Module} ($m_e$ blocks):
    \begin{equation}
    \begin{aligned}
        \text{FLOPs}_{{auf}} = & m_e(4 N_a^2 d + 24 N_a d^2) + \\ &m_e(4KN_ad+18Kd^2+4N_ad^2) .
    \end{aligned}
    \end{equation}
     where $K$ and $N_a$ are the number of slots and valid ads.
\end{enumerate}

Finally, the total Flops of EGA-V1 is calcalated as:
\begin{equation}
\begin{aligned}
     \text{FLOPs}_{{EGA-V1}}  =& \text{FLOPs}_{{gcf}} + \text{FLOPs}_{{mif}} + \text{FLOPs}_{{auf}}.
\end{aligned}
\end{equation}

Both the computational complexities of MCA and EGA-V1 are proportional to the number of candidates($N$). With $N_r = \alpha N$, $m_r = m$, and $m = 2m_c$, when $N$ is large, the ratio of proportions can be approximated as follows:

\[
\frac{\text{FLOPs}_{{EGA-V1}}}{\text{FLOPs}_{{mca}}} \approx \frac{2m_k N L d^2}{m_r N_r 4L^2 d}
\]

In our case, with $\alpha = 0.033$, this results in a ratio of $0.97$. The slight difference in proportions leads to a significant FLOPs gap as the number of candidates increases, demonstrating EGA-V1's capability and efficiency within our design. 



\subsection{Auction Constraints}
\label{sec:app_B}
To ensure incentive compatibility (IC) in our model, we adopt the concept of \textit{ex-post regret}~\cite{dutting2019optimal,zhu2024contextual} to quantify the potential gain an advertiser could obtain by untruthfully reporting their bid. This formulation enables us to enforce IC constraints in a differentiable manner, suitable for end-to-end optimization.

Formally, given the generated sequence $Y$, ad $x_i\in Y$ with true valuation $v_i$, the ex-post regret is defined as:
\begin{equation}
    \text{tgt}_i(v_i, Y) = \max_{b^\prime_i} \left\{ u_i(v_i; b^\prime_i, \bm{b}_{-i}, Y) - u_i(v_i; b_i, \bm{b}_{-i}, Y) \right\},
\end{equation}
where $b_i$ is the truthful bid, $b'_i$ is a potential misreport, and $\bm{b}_{-i}$ represents bids excluding the item $x_i$.
The IC constraint is satisfied if and only if $\text{rgt}_i = 0$ for all advertisers. In practice, we approximate this using $M$ sampled valuations from distribution $\mathbb{F}$, the empirical ex-post regret for ad $x_i$ is
\begin{equation}
    \widehat{\text{tgt}}_i = \frac{1}{M} \sum_{j=1}^{M} \text{tgt}_i(v_i^j, Y).
    \label{eq:empirical_regret}
\end{equation}
We then formulate the auction design problem as minimizing the expected negative revenue under the constraint that the empirical ex-post regret remains zero for each ad $x_i$:
\begin{equation}
    \min_{\bm{w}} \; - \mathbb{E}_{\bm{v} \sim \mathbb{F}} \left[ \text{Rev}^{\mathcal{M}} \right], \quad \text{s.t.} \quad \widehat{\text{tgt}}_i = 0, \; \forall i \in [N],
    \label{eq:ic_opt}
\end{equation}
With the optimization loss of payment network defined in Equation~\eqref{eq:pay_loss}, the proposed EGA-V1 ensures IC approximately.

\newpage
\bibliographystyle{ACM-Reference-Format}
\bibliography{ref}

\newpage
\end{document}